\documentclass[structabstract]{aa}  
\usepackage{graphicx}
\usepackage{natbib}
\usepackage{txfonts}
\newcommand{\eqn} [1] {

\begin{equation}
#1
\end{equation}}

\newcommand{\eqna} [1] {
\begin{eqnarray}
#1
\end{eqnarray}}
\newcommand{\eq} [1] {{Eq.~(\ref{#1})}}
\newcommand{\eqs} [1] {{Eqs.~(\ref{#1})}}

\begin{document}
%
   \title{Intrinsic photometric characterisation of stellar oscillations and granulation 
}

   \subtitle{Solar reference values and CoRoT response functions}

   \author{E. Michel\inst{1},
          R. Samadi\inst{1},
          F. Baudin\inst{2},
          C. Barban\inst{1},
          T. Appourchaux\inst{2},
          \and
          M. Auvergne\inst{1}
          }

   \institute{Observatoire de Paris-LESIA, CNRS (UMR~8109), Universit\'e Pierre et Marie Curie, Universit\'e
            Denis Diderot, pl. J. Janssen, F-92195 Meudon, France\\
         \and
             Institut d'Astrophysique Spatiale, UMR8617, Universit\'e Paris X, b\^at.121,
              F-91405 Orsay, France\\
             }

   \date{Received ...; accepted ...}

 
  \abstract
   { Measuring amplitudes of solar-like oscillations and the granulation power spectral density
   constitute two promising sources of information to improve our understanding and description
   of the convection in outer layers of stars. However, different instruments, using different
   techniques and different band passes, bring measurements which cannot be directly compared
   neither to each other nor to theoretical values.
   }
   { In this work, we define simple response
   functions to derive intrinsic oscillation amplitudes and granulation power density, from
   photometry measurements obtained with a specific instrument on a specific star.
   }
   {We test this method on different photometry data sets obtained on the Sun with two different
    instruments in three different band passes.
   }
   {We show that the results are in good agreement and we establish
   reference intrinsic values for the Sun in photometry. We also compute the response functions
   of the CoRoT instrument for a range of parameters representative of the Main Sequence 
   solar-like pulsators to
   be observed with CoRoT. We show that these response functions can
   be conveniently described by simple analytic functions of the effective temperature of the
   target star.
   }
   {}

   \titlerunning{Intrinsic photometric characterisation...} 

   \maketitle
%

\section{Introduction}

Solar-like oscillations are being detected in a rapidely growing
number of stars \citep[see e.g.][]{Bedding07}. The excitation of these oscillations first observed in the Sun is
attributed to the acoustic noise generated by convection in the outer layers of stars and the
measurement of their amplitude is a source of information on the
convection process \citep[see e.g.][]{Samadi07,Samadi07b}.  The
existing theoretical works generally bring parametric scaling laws 
calibrated on the Sun. However, as noticed by \citet{Kjeldsen05}, measurements made on
different stars with different instruments using different techniques in velocimetry or
photometry, in different spectral lines or band passes, have different sensitivity to the
oscillations. They cannot be compared directly to each other, nor to theoretical values. The
comparison to the Sun is not straightforward either, since the different existing data sets
obtained on the Sun have not been translated into a proper standard reference suited for
comparison with stars. \citet{Kjeldsen05} initiated such a normalization work and a
comparison between several stars. Then, very recently, \citet{Kjeldsen08} measured the
solar oscillation amplitude with stellar techniques, aiming at setting up a consistent reference
for stellar oscillation measurements.  This was done in velocimetry, since till now the vast
majority of solar-like oscillations measured in other stars has been obtained with this
technique. However, CoRoT \citep{Baglin06}  has started bringing photometric
measurements of oscillation in solar-like pulsators which will need to be measured
quantitatively and compared with those of the Sun and with those obtained in velocimetry.
In addition to oscillations, rapid photometry might allow measuring, in approximately the
same domain of frequency, the power density spectrum contribution associated with the
stellar granulation. Granulation being a manifestation of the convective motions at the
photosphere level, the profile of its power density spectrum is expected to reflect
characteristic time scales and geometrical scales associated with the convection process
as described by heavy 3D numerical simulations \citep[see e.g.][]{Ludwig06,Trampedach98} or by 
parametrized models \citep[see e.g.][]{Baudin06}.

In the present work, we consider measurements of solar photometrical variations obtained
with two different instruments in four different band passes (SOHO/VIRGO/PMO6 and SPM
three channels).
In the corresponding instrumental power density spectra, we fit contributions from the solar
background and from the acoustic oscillations (Sect.2).
Then, in Sect.3, 
we establish a simple instrumental response function
relating the instrumental power density measurement to the intrinsic bolometric luminosity
relative variation.
These response functions can be applied to infer intrinsic (bolometric) power density of solar
background from specific photometry measurements. They also can be used to derive intrinsic
amplitude of solar radial oscillations from the same data.  We discuss how they can be
adapted for non radial modes. Following \citet{Kjeldsen05} and \citet{Kjeldsen08}, we
also propose to relate the oscillation mean power density measurement to an intrinsic
amplitude chosen here to be the bolometric amplitude for radial modes.
In Sect.4, we show that the results obtained with the different data sets considered here 
are consistent to a
good approximation and allow us to produce a reference value of bolometric radial oscillation
amplitude for the Sun observed as a star, and a reference bolometric power density spectrum
for Solar granulation.
Then (Sect.5), we compute the response functions adapted to the CoRoT instrument for stars
representative of potential solar-like pulsators on the Main Sequence in terms of effective
temperatures, log g values and chemical compositions. We show that to a great extent, the
dependency with log g and chemical composition can be neglected and that the CoRoT
response functions can be conveniently described with a good precision by analytic functions
of $T_{\rm eff}$.
 

\section{Observational material and power density spectra}

   We consider four data sets obtained on the Sun with different techniques and different band
pass by SOHO/VIRGO/PMO6 (essentially bolometric variations) and by
SOHO/VIRGO/SPM (photon counting) in three narrow (5nm) bands at 402nm (blue), 500nm
(green) and 862nm (red) \citep{Frohlich97}.
For each of these time series, we compute the power density spectrum shown in Fig1 and
Fig2. Following the technique proposed by \citet{Kjeldsen05} for stellar oscillations
measurements, we smooth these spectra with a boxcar of width 405 $\mu$Hz corresponding to
3 times the solar large separation (135 $\mu$Hz). 

Then, we perform a least square fit of each spectrum with three
components: a flat white noise contribution essentially due to photon counting noise, the solar
background contribution detailed hereafter, and on top, the stellar oscillation spectrum
contribution.
For the solar background contribution, following \citet{Harvey85} and
\citet{Andersen98a},
we consider a sum of powerlaws: $P(\nu)= \Sigma_i P_i(\nu)$, and 
$P_i(\nu)=a_i {\zeta_i}^2 {\tau}_i / (1+ (2 \pi \tau_i \nu)^{C_i})$ (also noted 
$P_i(\nu)=A_i / (1+ (B_i \nu)^{C_i})$ for convenience hereafter) ,  with $\nu$ the frequency, 
$\tau_i$ the characteristic time scale and
$C_i$ the slope at high frequency associated with each powerlaw, and $a_i$ a normalizing
factor such as: ${\zeta_i}^2 = \int P_i(\nu)$ d$\nu$ corresponds to the variance of the corresponding
time series. Note that in the case of \citet{Harvey85}, $C_i$ being set to 2, $a_i=2$.
This corresponds to a signal which autocorrelation in time has a decreasing exponential
behaviour. However, as mentioned by \citet{Harvey85}, other values for decay rate power of
time might be found for different type of data probing the atmosphere at different heights
\citep[see e.g.][]{Andersen98a}.

The physical processes most commonly considered in the solar background and represented
by such powerlaws are: 
activity (predominant up to $\sim$10$\mu$Hz), supergranulation (up to $\sim$ 100$\mu$Hz), 
mesogranulation (up to $\sim$ 1 mHz), and granulation \citep[see e.g.][]{Andersen98b,Anklin98,Aigrain04}. 
In the present study, we will focus on
the two latter processes showing significant contribution above 100 $\mu$Hz, in the frequency
domain where oscillations are found.

An estimate of the two first contributions (white noise and solar background) is obtained by a
simultaneous fit of the spectrum outside the domain where the oscillation signal is seen
with function $D+ \Sigma_i P_i(\nu)$, where $D$ represents the white noise contribution. After
subtraction of these two components, we isolate the one due to stellar oscillations.

   \begin{figure}
   \centering
   \includegraphics[width=9cm]{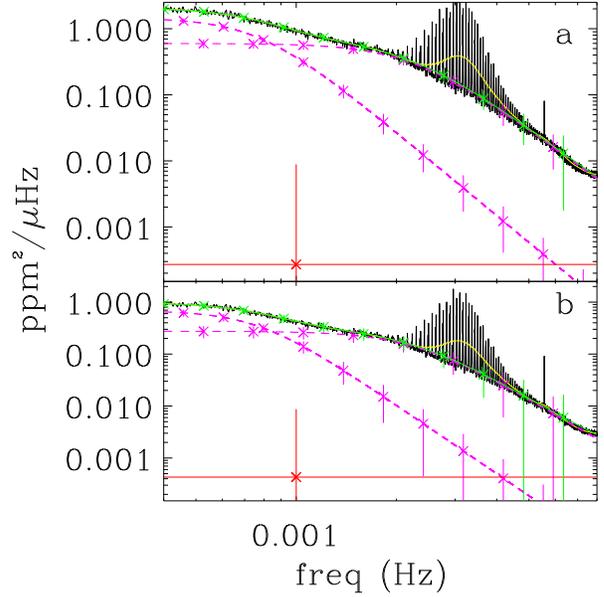}
   \caption{ Observational power density spectrum obtained for: 
SPM-blue ({\bf a}) 
and SPM-green ({\bf b}) data
over 700 days; a moving mean is applied with 
a 4 $\mathrm{\mu}$Hz boxcar (plain black line); 
the same
spectrum highly smoothed with a 0.405 mHz boxcar (3 $\Delta$) 
is superimposed (plain light grey line [yellow]);
Individual
powerlaws associated with granulation and mesogranulation 
are shown (dash lines [purple]);
The white noise component (horizontal line [red]); 
The global fit of solar background + white noise components is represented 
(plain grey line [green])
but differs from the mean power density only in the domain of oscillations. 
Vertical error bars associated with the fit precision are
illustrated at different frequencies for each component of the fit. For sake of clarity, in the case 
of the white noise component,
the error bar is represented only once at 1mHz. 
               }
              \label{Figinstpdsfitspmpmo6a}%
    \end{figure}

The two powerlaw components ( 7 parameters: $A_i$,$B_i$,$C_i$, and $D$) give satisfactory fit of the background for our purpose and we
do not find necessary to include other components like supergranulation or activity. 

As shown by error values
in Tab.~1, the fit give satisfactory results in the case of SPM data, especially for the blue
and green channels. In the case of SPM/red channel, the coefficients are obtained with very
large error bars and in the case of PMO6, the convergence precision is even worse, due to the
larger white noise component. We note that these fits all suggest a value $C_i$ around 4, in
agreement with the results obtained by \citet{Andersen98a}.
We thus decided to fit again the previous function, but
forcing the $C_i$ coefficients to the value 4, thus reducing the number of free parameters
to 5 and obtaining more precise determination of them.  

\begin{figure}
   \centering
   \includegraphics[width=9cm]{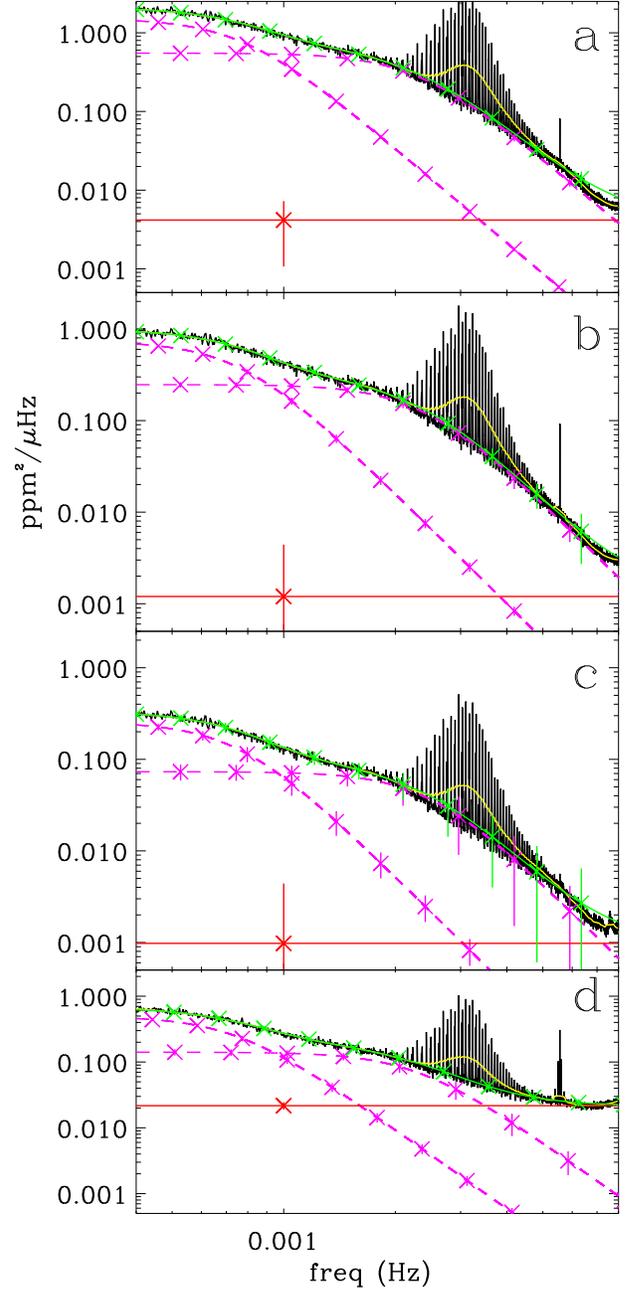}
   \caption{ Same as Fig.1 for, from top to bottom: SPM-blue({\bf a}),
    SPM-green({\bf b}), SPM-red({\bf c}) data
    over 700 days and PMO6({\bf d}) data over 800 days, 
    but here we forced $\mathrm{C_i}=4$. 
               }
              \label{Figinstpdsfitspmpmo6b}%
    \end{figure}

The results shown in Fig.~\ref{Figinstpdsfitspmpmo6b} (see also Tab.~2) are very satisfactory,
and we will refer to these values hereafter.
%
\begin{table}
\caption{Fit with seven parameters. Values of the parameters are given as well as the associated
one-sigma error estimates.}             
\label{table:1}      
\centering                          
\begin{tabular}{c c c c c c c c}        
\hline\hline                 
Data  & $A_1$ & $B_1$ & $A_2$ & $B_2$ & $D$ & $C_1$ & $C_2$  \\    
   & $\sigma_{A1}$ & $\sigma_{B1}$ &$\sigma_{A2}$ & $\sigma_{B2}$ &$\sigma_D$&$\sigma_{C1}$&$\sigma_{C2}$  \\    
  & (${{{ppm}^2}\over{\mu{\rm Hz}}}$) & (s) & (${{{ppm}^2}\over{\mu{\rm Hz}}}$) & (s) & (${{{ppm}^2}\over{\mu{\rm Hz}}}$) &  &   \\    
\hline                        
   SPMb & 1.46 & 1297 & 0.60 & 444 & 3 $10^{-4}$ & 4.2 & 3.7   \\      
        & 0.12 &   22 & 0.09 & 27  & 8.6 $10^{-3}$ & 0.4 & 0.5  \\      
   SPMg & 0.69 & 1300 & 0.28 & 438 & 4 $10^{-4}$ & 4.4 & 3.8 \\
        & 0.10 & 41   & 0.07 & 50 & 8.3 $10^{-3}$ & 0.8 & 1.1 \\
   SPMr & 0.23 & 1320 & 0.09& 438 & $-2$ $10^{-4}$ & 4.6 & 3.4  \\
        & 0.10 &  117 & 0.08& 185 & 1 $10^{-2}$ & 2.4 & 3.0  \\
   PMO6 & 0.54 & 1350& 0.13& 409 &  1.87 $10^{-2}$ & 3.6 & 3.8 \\
        & 0.20 & 110  & 0.12& 156 &  1.20 $10^{-2}$ & 1.6 & 3.0 \\
\hline                                   
\end{tabular}
\end{table}
%
%
\begin{table}
\caption{Fit with five parameters }             
\label{table:2}      
\centering                          
\begin{tabular}{c c c c c c}        
\hline\hline                 
Data set & $A_1$ & $B_1$ &$A_2$ & $B_2$ &$D$  \\    
         & $\sigma_{A1}$ & $\sigma_{B1}$ &$\sigma_{A2}$ & $\sigma_{B2}$ &$\sigma_D$  \\    
  & (${{{ppm}^2}\over{\mu{\rm Hz}}}$) & (s) & (${{{ppm}^2}\over{\mu{\rm Hz}}}$) & (s) & (${{{ppm}^2}\over{\mu{\rm Hz}}}$)   \\    
\hline                        
   SPMb & 1.52 & 1292 &0.55 & 433 & 4 $10^{-3}$   \\      
        & 0.02 &   18 &0.02 & 12 & 3 $10^{-3}$    \\      
   SPMg & 0.74 & 1302 &0.25 & 419 &  1 $10^{-3}$ \\
        & 0.02 & 37   &0.02 & 27 & 3 $10^{-3}$ \\
   SPMr & 0.26 & 1321 &0.07& 403 & 1 $10^{-3}$  \\
        & 0.02 &  105 &0.01& 89 & 3 $10^{-3}$  \\
   PMO6 & 0.50 & 1349& 0.14& 439 & 20 $10^{-3}$ \\
        & 0.02 & 55  & 0.02& 42 & 3 $10^{-3}$  \\
\hline                                   
\end{tabular}
\end{table}
%

\begin{figure}
   \centering
   \includegraphics[width=9cm]{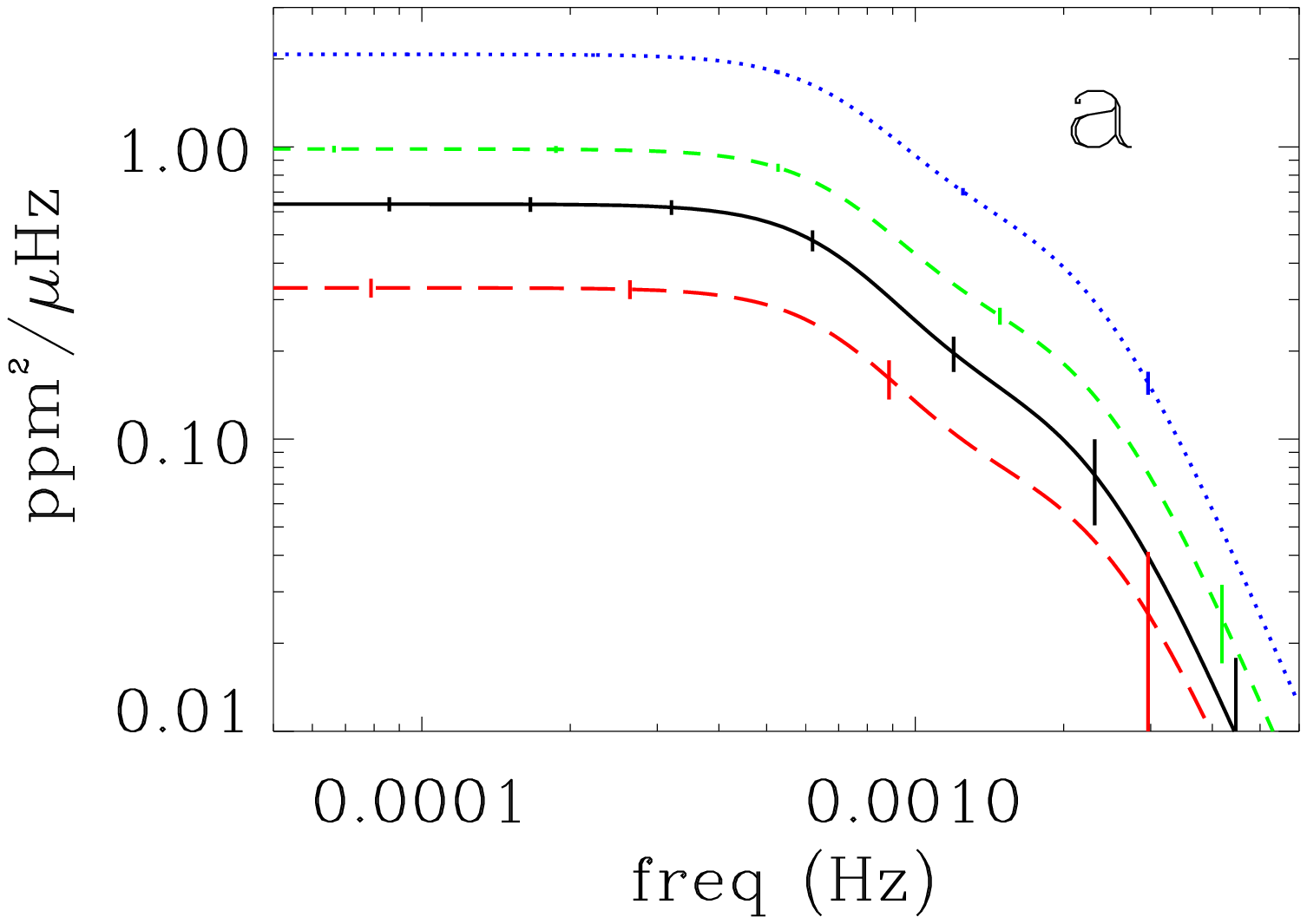}
   \includegraphics[width=9cm]{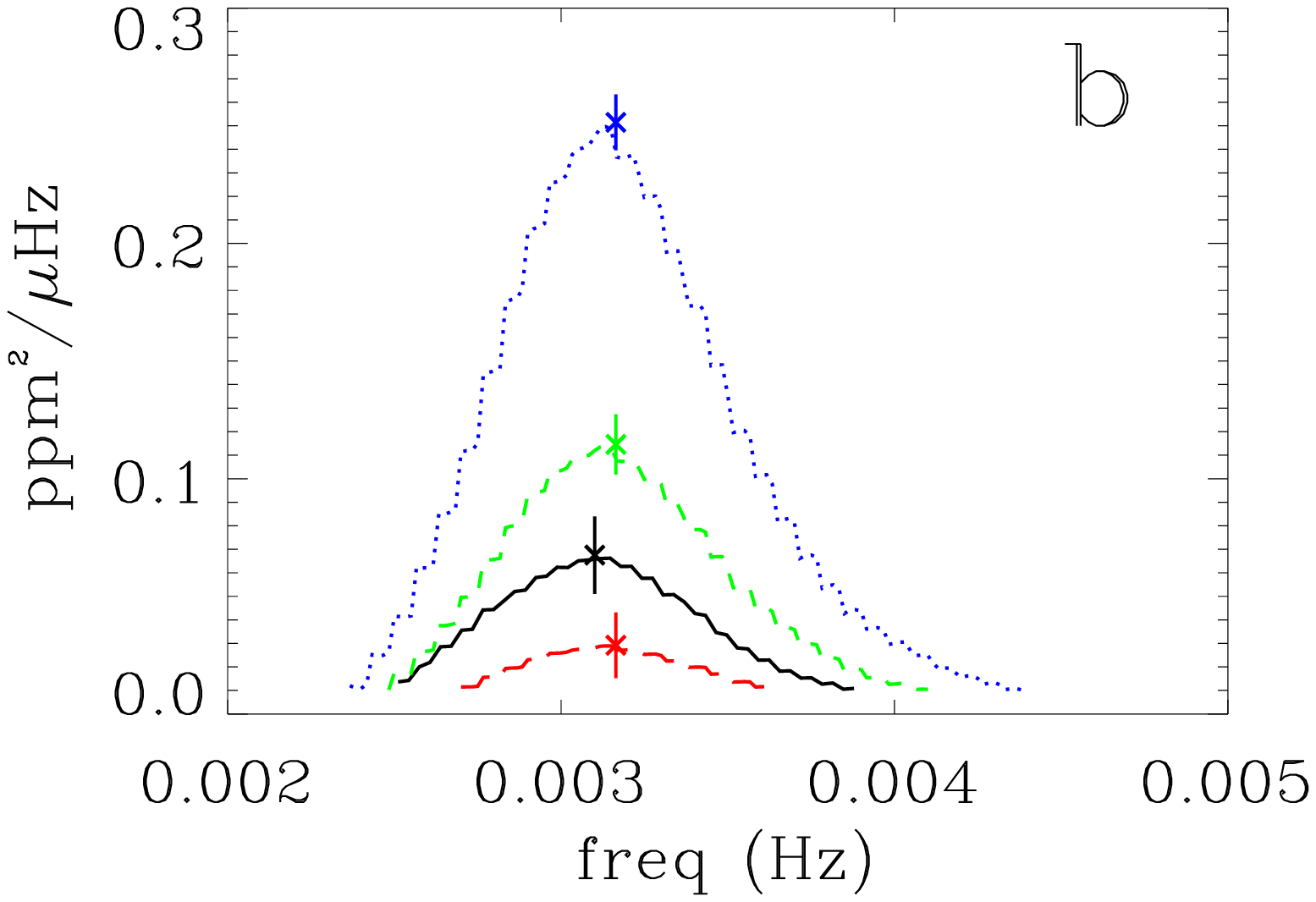}
   \caption{ {\bf a:} Observational instrumental power density 
spectrum associated with the stellar
background contribution and estimated as described in the text, 
for PMO6 data (plain black line), SPM-blue (dot [blue]), 
SPM-green (dash [green]), SPM-red (long dash [red]); 
{\bf b:} same for the oscillation contribution. 
               }
              \label{Figinstpds}%
    \end{figure}

 As could be expected, the level of the intensity components ($A_1$ and $A_2$) attributed to 
granulation and
mesogranulation is very different in the measurements associated to different techniques and
different band passes (see Fig.~\ref{Figinstpds} top). The same is true for the contribution associated with the
oscillations (Fig.~\ref{Figinstpds} bottom), stressing the necessity to establish a 
reference independent of the
instrument for the Sun oscillations and for comparison with other stars to be observed with
other instruments.

%

\section{Instrumental response functions}

In this section we establish a relation between
intensity variation observed with a given instrument 
(hereafter 'instrumental flux variation')
and an intrinsic quantity defined as the 'bolometric luminosity
variation'. This relation features a response function characteristic
of the instrument. 

We derive the response function for an individual non-radial mode (Sect.~3.2), then for
a smoothed power density spectrum (Sect.~3.3), and finally for the granulation (Sect.~3.4).

This is done taking into account both the band-width of the 
instrument, the spectral energy distribution of the given star
(approximated by a black body law) and the
dependence of the stellar limb-darkening with the wavelength
(given by stellar atmosphere models).

\subsection{Instrumental flux variation and local temperature variation}

Here, we express the relative instrumental 
flux variation $\delta I/\bar{I}$ as a function of 
the local relative variation 
of the temperature at the stellar photosphere $\delta T(\theta,\phi)/\bar{T}$.

First we define  
the relative variation of the instrumental flux $I$:
\eqn{
\left ( { {\delta I} \over \bar{I} } \right ) (t) = { { \int_\lambda d\lambda \, E(\lambda)
    \, \delta F_\lambda } \over { \int_\lambda d\lambda \, E(\lambda)
    \, F_\lambda }  }
\label{delta_Ia}
}
where $E(\lambda)$ is the global efficiency in terms of energy of the instrument 
at the wavelength $\lambda$, $F_\lambda$ the flux received from the star at the wavelegth
$\lambda$ and $\delta F_\lambda$ its variation. 

Then, following the approach of \citet{Berthomieu90}, we show 
(see Appendix~A) that
$F_\lambda$ and $\delta F_\lambda$ can be approximated as 
\eqn{
F_\lambda =  2 \pi \,  H_\lambda \, G_\lambda \, {B}_\lambda
\label{F_2a}
}
where
$B_\lambda$ is the black body law evaluated at the photosphere, i.e. at $T=T_{\rm
  eff}$ and where we have defined
\eqn{
 H_\lambda \equiv  2 \, \left (   \int_{-1}^{1}  d\mu \,
 g_\lambda(\mu)  \right) ^{-1}
\label{H_lambdaa}
}
and
\eqn{
G_\lambda \equiv     \int_0^1 d \mu \, \mu \,
 \, g_\lambda(\mu) \; .
\label{G_lambdaa}
}
with $g_\lambda$ the limb-darkening function, $\mu=cos(\theta)$ and
$\theta$ and $\phi$ refer to the spherical coordinates for a z-axis
pointing toward the observer (observer reference frame)

and
\eqn{
\delta F_\lambda =   H_\lambda \,  \left ( { {d B_\lambda} \over {d \ln T} } \right
) \,  \int_0^{2 \pi}  d \phi \, \int_0^1 d \mu \, \mu \,
 \, g_\lambda(\mu) \, \left ( { {\delta T} \over  \bar{T} } \right
)
\label{delta_F_3a}
}
where $T$ is the temperature, $\delta T$ its variation, the meaning of other 
terms staying the same. 

At this stage, we thus have expressed the instrumental 
flux relative variation $\delta I/\bar{I}$
as a function of $\delta T/\bar{T}$ the local ($\mu$,$\phi$) relative variation 
of the temperature at the photosphere.

\subsection{Response function for an individual non-radial mode}

Here, in the case of an individual oscillation mode, we show that we can
relate $\delta T(\theta,\phi)/\bar{T}$ with 
a 'bolometric luminosity relative variation' $(\delta L/\bar{L})_{\ell,m}$, 
defined as an extension of the specific case 
of radial modes where $(\delta L/\bar{L}) = 4 \delta T_{eff}/\bar{T}_{eff}$.

As detailed in Appendix~A, we consider the relative
temperature fluctuations associated with a mode with degree $\ell$ and
azimuthal order $m$:
\eqn{
\left ( { {\delta T} \over  \bar{T} } \right ) (t,\theta,\phi)  = \Theta_{\ell,m}(t) \,
  Y_\ell^m (\theta^\prime,\phi^\prime)
\label{dT_T_osc_2}
}
where $\Theta_{\ell,m}(t)$ is  the \emph{intrinsic} and
\emph{instantaneous} mode amplitude
in terms of temperature fluctuation,   $Y_\ell^m$ is the spherical
harmonic associated with the mode with degree $\ell$ and azimutal
order $m$, and $(r,\theta^\prime,\phi^\prime)$
(resp. $(r,\theta,\phi)$) the spherical coordinate system in
the pulsation frame (resp. in the observer frame).
As discussed in Appendix~A, $\left ( { {\delta T} /  \bar{T} } \right ) $ and hence $
\Theta_{\ell,m}(t)$ are evaluated at the photosphere.

For a radial mode, the \emph{bolometric} and \emph{instrinsic}
luminosity fluctuation is related to the relative instrinsic
temperature fluctuation as:
\eqn{
\left (\delta L \over  \bar{L} \right )_{\ell=0} (t) = 4 \, \left (
{\delta T_{\rm eff} \over T_{\rm eff}} \right )_0 =  4 \, \Theta_0 (t)
\label{amp_L_l0}
}
where $T_{\rm eff}$ is the effective temperature and $L$ the
luminosity of the star.
Then, by extension of the radial case, we define, in the general case,
the \emph{bolometric} and \emph{instrinsic} mode amplitude in terms of
luminosity  the quantity:
\eqn{
\left (\delta L \over  \bar{L} \right )_{\ell,m}  (t) \equiv 4 \, \Theta_{\ell,m}(t)
\label{amp_L_2}
}

Note that, in the present case, 
since the mode excitation is a random process, we rather
consider the \emph{rms} quantities
\eqn{
 \left (\delta L \over \bar{L} \right )_{\ell,m}^{\rm rms}
 \equiv  \sqrt{
  \, \overline{ \left  ( {  {\delta L } \over \bar{L} } \right )_{\ell,m}^2 (t)} }   = 4\,  \sqrt{
  \, \overline{ \Theta_{\ell,m}^2 (t)} }
\label{amp_L_rms}
}
The \emph{rms} label will however be generally omited in the following for conciseness
of the notations. 

Then, we establish the relation between
$\left ( { {\delta I} /  \bar{I} } \right )_{\ell,m,i}$ (the
observed relative intensity fluctuations due to a given mode $(\ell,m)$,
for a given inclination $i$) 
and
the instrinsic mode amplitude:
\eqn{
 \left ( { {\delta I} \over
  \bar{I} } \right )_{\ell,m,i}  =  R_{\ell,m,i}
\,   \Theta_{\ell,m} = { R_{\ell,m,i}
\over 4 } \, \left (\delta L \over \bar{L} \right )_{\ell,m}
\label{delta_I_osc_3}
}
with $R_{\ell,m,i}$ is the instrumental response function associated
with the mode with degree $\ell$ and azimuthal order $m$ and inclination $i$. 
The expression for $R_{\ell,m,i}$ is:
\eqn{
R_{\ell,m,i}  \equiv  { {\int_\lambda d\lambda \, E(\lambda)
    \,  { \displaystyle{{d B_\lambda} \over {d \ln T}} } \,  S_{\ell,m,i}  (\lambda)   } \over { \int_\lambda d\lambda \, E(\lambda)
    \, B_\lambda }   }
\label{R_lm_3}
}
where 
 $S_{\ell,m,i}(\lambda)$ is the so-called 'visibility' coefficient
 associated with the mode.

The visibility coefficient, $S_{\ell,m,i}$,
 measures the contribution of the mode integrated over the projected stellar surface, taking into account the
effect of the limb-darkening \citep[see
  e.g.][]{Dziembowski77,Berthomieu90}. Expression for $S_{\ell,m,i}$ is
given in \eq{S_l_m}. 
Note that in the case of radial modes, $S_0$ is independent of $\lambda$ and $S_0=1$ by definition.

An interesting property of the visibility coefficients $S_{\ell,m,i}$ makes that, assuming equipartition of
energy among different modes of a same multiplet, the global visibility contribution of each
multiplet (composed of modes of same radial order $n$, same degree $\ell$, and different azimuthal
order $m$) is independent of the inclinaison $i$ \citep[][]{Dziembowski77,Toutain93}. 
It is thus possible to compute a global visibility function $S_{\ell} =\sqrt{\Sigma_m S^2_{\ell,m,i}}$,
which is independent on $i$ for the different multiplets. This property will be usefull in the next section.


\subsection{Response function for smoothed oscillation power density spectrum}

In the case of stellar observations, as remarked by \citet{Kjeldsen05}  the measurement of
individual modes or even individual multiplets might be delicate and it might give more
precise results to estimate oscillation amplitudes from the smoothed oscillation power density
contribution as represented in Sect.1.
In this case, as suggested by \citet{Kjeldsen05}, the oscillation power density contribution
($P_{\rm osc}$, in $ppm^2/\mu$Hz)  is smoothed over typically three or four times the
large separation ($\Delta$); then, once multiplied by the estimate of the
large separation (in $\mu$Hz), it is representative of ${P_n}^I$ the total
power (in $ppm^2$) concentrated in all modes present in one large
separation (of order $n$). Accordingly, we have
\eqn{
\ {P_n}^I \simeq 2 \, P_{\rm osc} \, \Delta
\label{PI_Posc}
}
where $\Delta$ is the large separation and
the factor $2$ multiplying
$P_{\rm osc}$ is introduced to take into account the
power density spread in the negative part of the spectrum.

Let define $P_{\ell,m,i}$ as the observed power (in $ppm^2$) associated with a
mode $(\ell,m)$, with inclination $i$.
Assuming that $\Theta_{\ell,m}$ is the same for all the modes
that are within the same separation and using \eq{amp_L_l0}, it can easily shown that :
\eqn{
\ {P_n}^I=\sum_{l,m} {P_{\ell,m}}^I  = R_{\rm osc}^2  \,
\overline{ \Theta_{0}^2 (t)}
= {R_{\rm osc}^2  \over 16 } \,
\overline{ \left  ( {  {\delta L } \over \bar{L} } \right )_{0}^2 (t)} 
\label{sum_P_lm}
}
with
\eqna{
R_{\rm osc} &  \equiv & \sqrt { \sum_\ell R_\ell  }^2
\label{R_osc} \\
R_{\ell}  & \equiv  & { {\int_\lambda d\lambda \, E(\lambda)
    \,  { \displaystyle{{d B_\lambda} \over {d \ln T}} } \,  S_{\ell}  (\lambda)   } \over { \int_\lambda d\lambda \, E(\lambda)
    \, B_\lambda }   }
\label{R_l}
}
and $\overline{ \Theta_{0}^2 (t)}$ 
(resp. $\overline{ \left  ( {  {\delta L } \over \bar{L} } \right )_{0}^2 (t)}$ ) is the 
mean square value of 
$\Theta_0(t)$ (resp. $\left ( \delta L / L \right )_0(t)$) for a radial mode.
Note that in \eq{sum_P_lm} the sum over $(\ell,m)$ is performed among all the  multiplets within the same
separation. The expression for the visibility coefficient $S_\ell$ is given
by \eq{S_l}.
The visibility factor associated with modes $\ell > 4$ can be
neglected. Accordingly, \eq{R_osc} can be simplified as:
\eqn{
R_{\rm osc} = \sqrt{ R_0^2 + R_1^2 + R_2^2 + R_3^2 }
\label{R_osc_2}
}
In practice, let consider $P_{\rm osc}$, the smoothed power density component
associated with oscillations derived from observations, as in
Sect.1. Using \eqs{PI_Posc}, \ref{sum_P_lm} and (\ref{amp_L_rms}), one obtains 
the  (\emph{rms}) bolometric amplitude normalised to
radial mode given by
\eqn{
A_{\rm bol,\ell=0} \equiv \left (\delta L \over \bar{L} \right
)_0^{\rm rms} = { 4 \over R_{\rm osc} } \, \sqrt{\displaystyle{2 \, P_{\rm osc} \, \Delta}} 
}
where $R_{\rm osc}$ is the
response function given by \eq{R_osc_2} and
computed for each data set using \eqs{R_l} and (\ref{S_l}). 

In the present work, the $S_\ell(\lambda)$ coefficients (\eq{S_l}) are computed,
taking into account monochromatic specific intensities derived from stellar atmosphere
models \citep[see][]{Barban03} with relevant $T_{\rm eff}$, [Fe/H],
and $\log g$.


\subsection{Response function for granulation}

As detailed in Appendix~A, since we are interested in \emph{rms} values with time and
assuming that these values are identical all over the stellar surface,
the granulation component can be treated in a
similar way than a radial mode.
Accordingly, the relation between the observed relative intensity
fluctuations  and the associated  intrinsic fluctuations  is
\eqn{
\left ( { {\delta I} \over \bar{I} } \right )_g (t)   =   R_g \,
\Theta_g (t) = { R_{g} \over 4 } \, \left (\delta L \over
\bar{L} \right )_g (t)
\label{delta_I_g_3}
}
where the quantities have the same meaning than previously for
radial modes but
subscript $g$ refers to the granulation and
\eqn{
R_{g}   =   R_{\ell=0,m=0}   =   { {\int_\lambda d\lambda \, E(\lambda)
    \,  { {d B_\lambda} \over {d \ln T} }   } \over { \int_\lambda d\lambda \, E(\lambda)
    \, B_\lambda }   }
}

As for the radial modes, we define the \emph{rms} and instrinsic
relative luminosity fluctuation due to granulation  the quantity
\eqn{
 \left (\delta L \over \bar{L} \right )_g^{\rm rms}
 \equiv  \sqrt{
  \, \overline{ \left  ( {  {\delta L } \over L } \right )_g^2 (t)} }    = 4\,  \sqrt{
  \, \overline{ \Theta_g^2 (t)} }  = { 4 \over R_{g} } \, \left (\delta I \over
\bar{I} \right )_g^{\rm rms} (t)
\label{amp_L_g_rms}
}

If we consider the power density contribution associated with granulation
($P_g$) determined in Sect.1, we can derive the corresponding
bolometric power density spectrum according to
\eqn{
P_{g,{\rm bol}}=16~P_g/{R_g}^2
}
which is expected to characterize granulation independently of the instrument considered.
The application to the different data sets obtained on the Sun ($R_g$ values are given in Tab.~3)
with different instrumental
techniques and with different band passes reveal a good agreement (see Sect.~4).

%
\begin{table}
\caption{Response functions for different sets of solar data }             
\label{table:3}      
\centering                          
\begin{tabular}{c c c c c}        
\hline\hline                 
Resp. Func. & ${\rm SPM_b}$ & ${\rm SPM_g}$ & ${\rm SPM_r}$ & ${\rm PMO6}$  \\    
\hline                        
   $R_{osc}(T_{\rm eff,Sun})$ & 11.63 & 9.02 & 5.26 & 7.15  \\
   $R_g(T_{\rm eff,Sun})$ & 6.24 & 5.02 & 3.06 & 4.00    \\      
\hline                                   
\end{tabular}
\end{table}
%


\section{Results for different data sets}


\subsection{A reference solar bolometric oscillation amplitude}

\begin{figure}
   \centering
   \includegraphics[width=9cm]{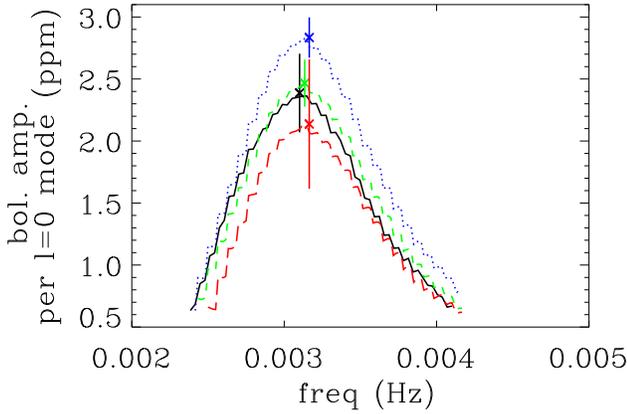}
   \caption{
Observational bolometric amplitude per radial mode estimated as described in the text, for
PMO6 data (plain black line), SPM-blue (dot [blue]), SPM-green (dash [green]),
SPM-red (long dash [red]). Error bars are given for the
estimate of the maximum (boxcar: 3 times Large Separation taken as 135 $\mu$Hz).
               }
              \label{Figbolamp}%
    \end{figure}

 The resulting estimates of the bolometric amplitude per radial mode are shown 
in Fig.~\ref{Figbolamp} ($R_{osc}$ values computed for the different data sets considered here are
given in Tab.~3).
We compare the curves obtained for each data set, with a special attention to the value at
maximum often taken as a convenient characteristic measurement of the oscillations
amplitudes in stars (see also Tab.~4). Although some residual of the initial difference 
seems to subsist (suggesting that our response function might be refined further), 
we notice a reasonable agreement of the different curves, within one-sigma error bar estimates.
This allows us to propose as reference for the Sun a $2.53 \pm 0.11$ppm of maximum
bolometric amplitude per radial mode (mean of the four values ponderated by $1/\sigma_i$).
We checked that this result was not affected significantly by changing the smoothing
boxcar width from 2 times to 4 times $\Delta$.

%
\begin{table}
\caption{Bolometric parameters. The last line corresponds to reference values resulting from
a mean of the values given in the other lines, ponderated by $1/{\sigma}_i$.}             
\label{table:4}      
\centering                          
\begin{tabular}{c c c c c c c c}        
\hline\hline                 
Data & $A_{1,bol}$ & $\zeta_1$& $\tau_1$ &$A_{2,bol}$ &$\zeta_2$ & $\tau_2$&$A_{bol,\ell=0}$  \\    
         & $\sigma_{A1bol}$ &$\sigma_{\zeta 1}$ & $\sigma_{\tau 1}$ &$\sigma_{A2bol}$ &$\sigma_{\zeta 2}$ & $\sigma_{\tau 2}$&$\sigma_{Abol,\ell=0}$ \\    
set & (${{{ppm}^2}\over{\mu{\rm Hz}}}$) & & (s) & (${{{ppm}^2}\over{\mu{\rm Hz}}}$) & & (s) & ($ppm$)   \\    
\hline                        
   SPMb & 0.62 & 8.2~$10^{-3}$& 206 &0.23 & 8.5~$10^{-3}$& 69&2.83  \\     
        & 0.01 & 1.~$10^{-4}$ &   3 &0.01 & 2.~$10^{-4}$& 2& 0.16  \\      
   SPMg & 0.47 & 7.1~$10^{-3}$& 207 &0.16 &7.2~$10^{-3}$ & 67& 2.47  \\
        & 0.01 & 1.~$10^{-4}$ & 6 &0.01 &  3.~$10^{-4}$ & 4& 0.19  \\
   SPMr & 0.44 & 6.8~$10^{-3}$ & 210 &0.13& 6.6~$10^{-3}$& 64& 2.14   \\
        & 0.03 & 4.~$10^{-4}$&  17 &0.02& 1.0~$10^{-3}$ & 14& 0.52   \\
   PMO6 & 0.50 &7.2~$10^{-3}$ & 215& 0.14&6.7~$10^{-3}$ & 70& 2.36  \\
        & 0.02 & 2.~$10^{-4}$& 9  & 0.02& 5.~$10^{-4}$& 7& 0.23   \\
\hline                        
   Ref & 0.52 &7.6~$10^{-3}$ & 208& 0.18&7.6~$10^{-3}$ & 68& 2.53  \\
        & 0.01 & 1~$10^{-4}$& 3  & 0.01& 2.~$10^{-4}$& 2& 0.11   \\
\hline                                   
\end{tabular}
\end{table}
%

\subsection{A reference bolometric granulation power density spectrum}

\begin{figure}
   \centering
   \includegraphics[width=9cm]{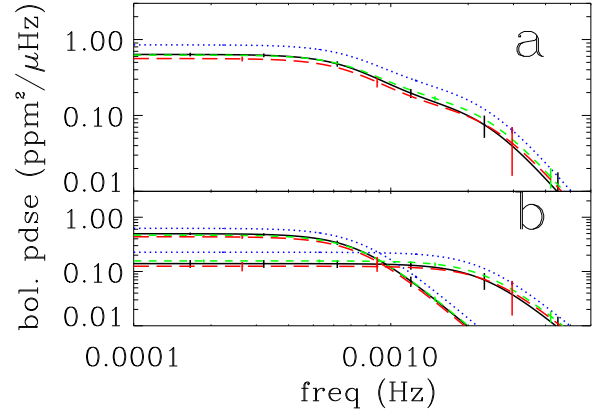}
   \caption{
Observational bolometric power density spectrum estimated as described in the text, for
PMO6 data (plain black line), SPM-blue (dot [blue]), SPM-green (dash [green]), 
SPM-red (long dash [red]). {\bf a:} granulation+
mesogranulation components; 
{\bf b:} granulation and mesogranulation individual components.
               }
              \label{Figbolgran}%
    \end{figure}

The different mean profiles of bolometric background power density spectra are shown in
Fig.~\ref{Figbolgran}. Here again, we notice the good agreement of the different curves.
Coefficients characterizing the different curves are given in Tab.~4 as well as
reference values proposed for the Sun background contribution.
Here again, the influence of the size of the smoothing boxcar (between 0.1 to 4 times $\Delta$) 
has been tested and 
found negligible within the present error bars.

\section{Response functions of CoRoT for objects on the Main Sequence}

\begin{figure}
   \centering
   \includegraphics[width=9cm]{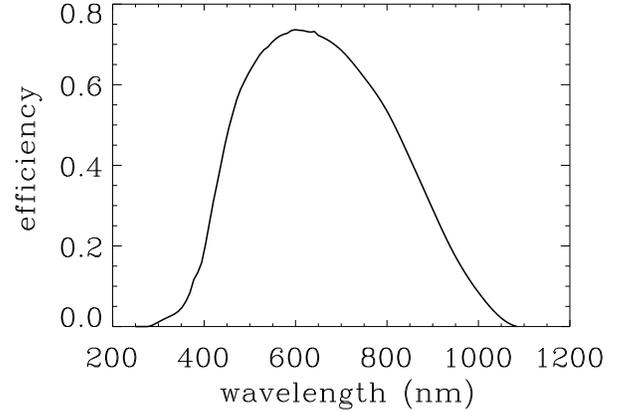}
   \caption{
CoRoT total efficiency.
               }
              \label{Coroteff}%
    \end{figure}

Stellar atmosphere models are computed with the Atlas~9 code \citep{Kurucz93}
in a modified version including the CGM convection \citep[see]{Heiter02}.
 Considering the CoRoT total efficiency shown in Fig.~\ref{Coroteff},  
we compute the CoRoT response functions for stellar atmosphere
models characterized by different values of $T_{\rm eff}$, $\log g$ and chemical compositions illustrative
of possible solar-like candidates on the Main Sequence ($-1<[Fe/H]<+1$, $3.9<\log g<4.5$,
$5800 < T_{\rm eff} <6750 K$).

   \begin{figure}
   \centering
   \includegraphics[width=9cm]{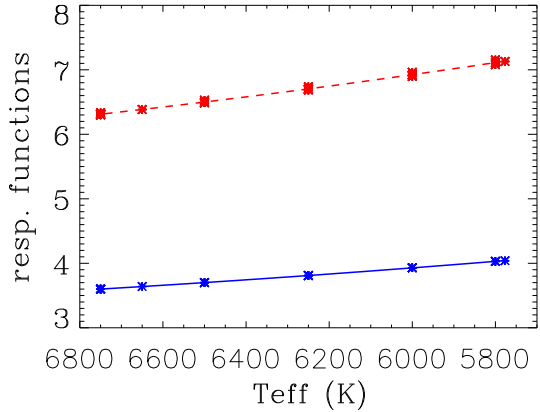}
   \caption{
Polynomial fit of the CoRoT response function $R_{osc}$ (dash [red]) 
and $R_g$ (plain [blue]) for different stellar atmosphere models
mentioned in the text.
               }
              \label{fitcorotcr}%
    \end{figure}

As shown in Fig.~\ref{fitcorotcr}, the dependency of the CoRoT response function $R_{osc}$ to $\log g$ and
chemical composition is small in the considered range. To a great extent (within $0.6$\%), it can be neglected
and $R_{osc}$ as $R_g$ can be described as simple polynomial functions of $T_{\rm eff}$ only:
$$R_{osc}(T_{\rm eff})= R_{osc}(T_{\rm eff,Sun})+A~(T_{\rm eff}-T_{\rm eff,Sun})+B~(T_{\rm eff}-T_{\rm eff,Sun})^2.$$
We proceed the same way for $R_g$.

Values of the parameters obtained for the fit are given in Tab.~5.
%
\begin{table}
\caption{Fit of CoRoT response functions }             
\label{table:5}      
\centering                          
\begin{tabular}{c c c c c}        
\hline\hline                 
Resp. Func. & $R(T_{\rm eff,Sun})$ & $A (K^{-1})$ & $B (K^{-2})$ & ${\chi}^2$  \\    
 & $\sigma_{R(Teff,Sun})$ & $\sigma_A$ & $\sigma_B$ &   \\    
\hline                        
   $R_{osc}(T_{\rm eff})$ & 7.134 & $-96.8~10^{-5}$ & $13~10^{-8}$ & $8~10^{-3}$    \\      
        & $9~10^{-3}$ & $4.4~10^{-5}$ &$4~10^{-8}$ &    \\      
   $R_g(T_{\rm eff})$ & 4.0420 & $-523~10^{-6}$ & $71~ 10^{-9}$ & $1.3~10^{-5}$   \\
        & $4~10^{-4}$  & $2~10^{-6}$    & $2~10^{-9}$ & \\
\hline                                   
\end{tabular}
\end{table}
%


\section{Conclusions}
Measurement of stellar oscillations or granulation brings instrumental values which depend
on the instrumental technique and bandpass and on the star considered. 
In this work, with the pourpose of helping future comparisons between 
stars observed in photometry,
   \begin{enumerate}
      \item we propose a simple expression for response functions connecting 
specific instrumental photometric measurements with intrinsic bolometric 
values for oscillation amplitudes and granulation power density. 
      \item we test and validate this expression on four sets of data obtained 
on the Sun, in four different band passes and with two different instrumental techniques. 
      \item we establish reference bolometric measurements for the Solar oscillation amplitudes
($ 2.53 \pm 0.11$ ppm) and for the Solar granulation power density. 
      \item we compute the response functions for the CoRoT instrument and give an analytic expression
for it. 
   \end{enumerate}

\begin{acknowledgements}
     SOHO is a mission of international collaboration between ESA and NASA. 
\end{acknowledgements}

\bibliographystyle{aa}


\begin{appendix} 

\section{Derivation of the instrumental response functions}

We derive here the relation between the \emph{observed}  flux
fluctuation and the \emph{intrinsic} temperature fluctuations
induced by the presence of non-radial modes or granulation on the
surface of the star. We proceed in the
manner of \citet{Berthomieu90}.  We summarize the main
steps and  emphasize the approximations that we adopt in the present study.
The flux, $F_\lambda$,  received from the star at the wavelength $\lambda$ is given by
\citep[see][]{Berthomieu90}:
\eqn{
F_\lambda = \int_{\cal A} d {\cal A} \, I_\lambda (\tau=0, \mu)
\label{F}
}
where ${\cal A}$ is the total oberved surface, $d {\cal A} = d \vec A
. \vec n$ the elementary observed surface around the direction of the observer,
$\vec n$ a unit vector in the direction of the observer,
$d \vec A$ the differential surface element perpendicular to the
stellar surface, $\tau$ the optical depth, $\mu=\cos(\theta)$, and
$I_\lambda(\mu)$ the specific  intensity at the wavelength
$\lambda$. We adopt a spherical coordinate system with the $z$-axis pointing
toward the observer. The specific intensity is assumed to be invariant
with respect to any rotation along the $z$-axis, this is why here
$I_\lambda$ depends only on $\mu$. Note that the integral of \eq{F} is evaluated
at the optical depth $\tau=0$.

We define the limb-darkening function, $g_\lambda$, as
\eqn{
g_\lambda \equiv { I_\lambda(\mu) \over I_\lambda(1) }
\label{limb}
}
where $I_\lambda(1) \equiv I_\lambda(\mu=1)$.
Finally, we define the mean intensity as the quantity
\eqn{
J_\lambda \equiv {1 \over {4 \pi}} \, \int d\Omega \, I_\lambda(\mu)
\label{J}
}
where $d \Omega$ is the elementary solid angle.
Using \eqs{limb} and (\ref{J}) we then derive the relation:
\eqn{
J_\lambda  =  I_\lambda(1) /  H_\lambda
\label{J_2}
}
where we have defined
\eqn{
 H_\lambda \equiv  2 \, \left (   \int_{-1}^{1}  d\mu \,
 g_\lambda(\mu)  \right) ^{-1}
\label{H_lambda}
}

According to \eqs{F}, (\ref{limb}) and (\ref{J_2}), a small
variation of $F_\lambda$  is given by
\eqna{
\delta F_\lambda & = & \int_{\cal A} \,  d {\cal A} \, \left ( \bar
       {g}_\lambda \, \bar{H}_\lambda  \delta J_\lambda
 \, +  \bar {J}_\lambda
 \, \bar{H}_\lambda \, \delta  g_\lambda +  \bar {J}_\lambda
  \, \bar{g}_\lambda  \, \delta H_\lambda \right )\nonumber  \\ &
  & + \, \delta  \left ( d {\cal A} \right ) \, \bar{J}_\lambda
 \, \bar{g}_\lambda
\label{delta_F}
}
where $\overline{()}$ refers to equilibrium quantity.
The first term in RHS of \eq{delta_F} corresponds to the perturbations of
the mean intensity evaluated at an effective optical depth $\tau=
\tau_0$ in the atmosphere.
This effective optical depth corresponds to the layer that contribute
predominantly to the variation of the emergent flux \citep[see][]{Berthomieu90}.
As in \citet{Berthomieu90}, we assume that $g_\lambda$ and hence
$H_\lambda$  do not depend on $\tau_0$ and are evaluated at $\tau=0$.

The three last terms in RHS of \eq{delta_F}
are the perturbation of the limb-darkening function and the
surface distortion \citep[for details see][]{Berthomieu90} .
All these perturbuations are shown to have a negligible
contribution to $ \delta F_\lambda$ compared to that of  $\delta
J_\lambda$.
Accordingly, \eq{delta_F} can be simplified as:
\eqn{
\delta F_\lambda = \int_0^{2 \pi}  d \phi \, \int_0^1 d \mu \, \mu  \,
g_\lambda(\mu) \, H_\lambda \, \delta J_\lambda  \;,
\label{delta_F_2}
}
where we have dropped $\overline{()}$
from $g_\lambda$ and $H_\lambda$.

 We place ourself in
Local Thermodynamic Equilibrium  and  assume adiabatic perturbations for linearisation, accordingly $\delta J_\lambda = \delta
B_\lambda$ where $B_\lambda$ is the black body law which
expression is
\eqn{
B_\lambda = { {2 h\, c^2 } \over \lambda^5} \, { 1 \over { e^{ {h\,c }
\, / \, {\lambda \, k \, T} } - 1 } }
}
where $T$ is the local temperature, $c$ the speed of the light, $h$
Planck's constant, and $k$ Boltzmann's constant.

The local variation of $B_\lambda$ is induced by a local variation of
$T$. Assuming small perturbations, we have
\eqn{
\delta J_\lambda = \delta B_\lambda = \left ( { {d B_\lambda} \over {d \ln T} }
\right ) \,
\left ( { {\delta T} \over  T } \right ) (\tau_0,t,\theta,\phi)
\label{delta_B}
}

Using \eqs{delta_B} and (\ref{dT_T_osc}), \eq{delta_F_2} can then be
written as:
\eqn{
\delta F_\lambda =   H_\lambda \,  \left ( { {d B_\lambda} \over {d \ln T} } \right
) \,  \int_0^{2 \pi}  d \phi \, \int_0^1 d \mu \, \mu \,
 \, g_\lambda(\mu) \, \left ( { {\delta T} \over  T } \right
)
\label{delta_F_3}
}

Finally, we approximate \eq{F} as
\eqn{
F_\lambda =  2 \pi \,  H_\lambda \, G_\lambda \, {B}_\lambda
\label{F_2}
}
where $B_\lambda$ is evaluated at the photosphere, i.e. at $T=T_{\rm
  eff}$ and where we have defined
\eqn{
G_\lambda \equiv     \int_0^1 d \mu \, \mu \,
 \, g_\lambda(\mu) \; .
\label{G_lambda}
}

The relative variation of the total flux $I$ received by the
instrument is finally given by
\eqn{
\left ( { {\delta I} \over \bar{I} } \right ) (t) = { { \int_\lambda d\lambda \, E(\lambda)
    \, \delta F_\lambda } \over { \int_\lambda d\lambda \, E(\lambda)
    \, F_\lambda }  }
\label{delta_I}
}
where $E(\lambda)$ is the global efficiency in terms of energy of the instrument at a
given wavelength. The function $E(\lambda)$ is normalised as
\eqn{
\int_0^{ +\infty } d \lambda\, E(\lambda) = 1
}

\subsection{Non radial oscillations}

In the case of a non-radial spheroidal mode, $ {\delta T} /  T$ is
by definition:
\eqn{
\left ( { {\delta T} \over  T } \right ) (\tau_0,t,\theta,\phi)  = \Theta_\ell(t,\tau_0) \,
  Y_\ell^m (\theta^\prime,\phi^\prime)
\label{dT_T_osc}
}
where $\Theta_\ell(t,\tau_0)$ is  the \emph{intrinsic} and
\emph{instantaneous} mode amplitude
in terms of temperature fluctuation,   $Y_\ell^m$ is the spherical
harmonic associated with the mode with a degree $\ell$ and azimutal
order $m$,  and $(r,\theta^\prime,\phi^\prime)$ the spherical coordinate system in
the pulsation frame.
The pulsation frame is choosen such that its
polar axis coincides with the star rotation axis.
The  spherical harmonic, $Y_\ell^m$, is here normalized as:
\begin{equation}
\int d\Omega^\prime \,  \left \| Y_{\ell}^m (\theta^\prime,\phi^\prime) \right \|^2
  = 4 \pi
\label{othogonalite}
\end{equation}
where $\Omega^\prime$ is the elementary solid angle  associated with
the pulsation coordinate system.
Note that for low $\ell$ degree, $ \Theta(t,\tau_0) $ is expected to negligibly depend on
$\ell$ ~\citep{Belkacem08}.

As shown by \citet{Berthomieu90}, for low $\ell$ degree, $\tau_0$ marginally
depends on $\ell$.
Furthermore, they show that --~ in the Sun ~-- the optical depth
$\tau_0$ is very close the the photosphere, which by definition corresponds to the
layer $T = T_{\rm   eff}$ and $\tau=2/3$.
Then, from here, we will assume that $\tau_0$ coincides with the
photosphere ($\tau=2/3$).

Using  \eqs{delta_F_3}, (\ref{F_2}), (\ref{delta_I}),  (\ref{G_lambda}), and
(\ref{dT_T_osc}), we
then derive the flux variation due to the mode:
\eqn{
\left ( { {\delta I} \over \bar{I} } \right ) (t) = R_{\ell,m,i}
\,   \Theta_\ell (t)
\label{delta_I_osc}
}
with
\eqn{
R_{\ell,m,i}  \equiv  { {\int_\lambda d\lambda \, E(\lambda)
    \,  { \displaystyle{{d B_\lambda} \over {d \ln T}} } \, G_\lambda
    \, H_\lambda \, S_{\ell,m,i}  (\lambda)   } \over { \int_\lambda d\lambda \, E(\lambda)
    \, B_\lambda \, G_\lambda
    \, H_\lambda}   }
\label{R_lm}
}
where we have defined the 'visibility' coefficient,
$S_{\ell,m,i}$, as the quantity:
\eqn{
S_{\ell,m,i}  (\lambda)  \equiv  { {  \int_0^{2 \pi}  d \phi \, \int_0^1 d \mu \, \mu \,
 \, g_\lambda(\mu) \, Y_\ell^m (\theta^\prime,\phi^\prime) } \over  {
  2 \, \pi \, \int_0^1 d \mu \, \mu \,
 \, g_\lambda(\mu)   } }
\label{S_l_m}
}
Note that, by definition of $S_{\ell,m,i}$, we have for $S_0$=1 a radial
mode.

By using stellar atmosphere models, we find
that --~ in the domain of  $T_{\rm eff}$ and gravity we are interested
here ~-- $G_\lambda \, H_\lambda$
varies slowly with $\lambda$ compared to $B_\lambda$ and $dB_\lambda\,
/ \,d\ln T$. Accordingly, \eq{R_lm} can be simplified as:
\eqn{
R_{\ell,m,i}  \equiv  { {\int_\lambda d\lambda \, E(\lambda)
    \,  { \displaystyle{{d B_\lambda} \over {d \ln T}} } \,  S_{\ell,m,i}  (\lambda)   } \over { \int_\lambda d\lambda \, E(\lambda)
    \, B_\lambda }   }
\label{R_lm_2}
}

Following  \citet{Dziembowski77}, we can  decompose $S_{\ell,m,i} (\lambda) $ as:
\eqn{
 S_{\ell,m,i} = q_{\ell,m}(i) \, S_{\ell}
}
with
\eqna{
S_{\ell} (\lambda) & = &  { {\int_0^{1} d\mu \,
  \,  \mu \,  g_{\lambda}(\mu) \, Y_\ell^0 (\mu) } \over {\int_0^1 d \mu \, \mu \,
 \, g_\lambda(\mu)  }
  } \label{S_l}\\
q_{\ell,m}(i) & =  & \sqrt{{(l-m) ! } \over {(l+m) !} } \,  \left |
P_{\ell}^{|m|} \right | \,  \cos(i)
}
where $i$ is the angle between the observer and the rotation axis and
$ P_{\ell}^{|m|}$ the associated Legendre function.

The bolometric flux variation, $\left ( { {\delta I} / \bar{I} } \right
)^{\rm bol}$, is obtained from \eq{delta_I_osc} by assuming in \eq{R_lm_2}
a constant $E(\lambda)$, this gives
\eqn{
\left ( { {\delta I} \over \bar{I} } \right )^{\rm bol} (t) = R_{\ell,m,i}^{\rm bol}
\,   \Theta_\ell  (t)
\label{delta_I_osc_bol}
}
with
\eqn{
R_{\ell,m,i}^{\rm bol}  \equiv  { \pi \over {\sigma \, T_{\rm eff}^4} } \, \int_\lambda d\lambda \,
    \,  { {d B_\lambda} \over {d \ln T} } \, S_{\ell,m,i}  (\lambda)
\label{R_lm_bol}
}
For a radial mode, $S_{0,0}=1$  and  $R_{{\rm
    bol},0,0}= 4$. We have then  for a radial mode:
\eqn{
\left ( { {\delta I} \over \bar{I} } \right )^{\rm bol} (t) = 4 \,
\Theta_0 (t)
\label{delta_I_bol}
}

By definition of the effective temperature ($T_{\rm eff}$) and the stellar
radius $R_*$, the total luminosity of the star, $L$, is given by the Steffan's
law:
\eqn{
L = 4 \pi \, \sigma \, T_{\rm eff}^4 \, R_*^2 \;
\label{L}
}
where $\sigma$ is Steffan's constant.
Variation of the stellar radius due to the mode can be neglected.
Accordingly, the relative variation of $L$  due to a radial
mode is given by the relation
\eqn{
\left (\delta L \over \bar{L} \right ) = 4 \left
(\delta T_{\rm eff} \over {\bar{T}_{\rm eff}} \right )
\label{amp_L_0}
}
Beside, we have again for a radial mode:
\eqn{
\left ( { {\delta I} \over \bar{I} } \right )^{\rm bol}  = \left ( { {\delta L} \over \bar{L} } \right )
\label{delta_I_bol_L}
}

Then, according to \eqs{delta_I_bol} and (\ref{amp_L_0}), we have
\eqn{
\left ({ \delta T_{\rm eff} \over { \bar{T}_{\rm eff}} } \right )   =  \Theta_0
}

As a conclusion, for a radial mode, $\Theta_0$ (resp. $\left ( {
  {\delta I} / \bar{I} } \right )^{\rm bol}$) is then directly related to a variation
of $T_{\rm eff}$ (resp. $L$).
On the other hand, for a \emph{non-radial} mode,  $\left ( { {\delta I} / \bar{I} } \right
  )^{\rm bol}$  is related to the instrinsic mode amplitude in terms
  of temperature, $\Theta_\ell $, through the coefficient given by
  \eq{R_lm_bol} that depends
  on the mode geometry and the limb-darkening law.
However, by extension of the radial case, we define, in the general case,
the \emph{bolometric} and \emph{instrinsic} mode amplitude in terms of
luminosity  the quantity:
\eqn{
\left (\delta L \over  \bar{L} \right )_\ell  \equiv 4 \, \Theta_\ell
\label{amp_L}
}

Now, according to \eqs{delta_I_osc} and (\ref{amp_L}),  we can  writte:
\eqn{
 \left ( { {\delta I} \over
  \bar{I} } \right ) (t) =  R_{\ell,m,i}
\,   \Theta_\ell = { R_{\ell,m,i}
\over 4 } \, \left (\delta L \over \bar{L} \right )_\ell
\label{delta_I_osc_2}
}
The \eq{delta_I_osc_2} then relates the observed intensity
fluctuations to the bolometric and instrinsic mode amplitude in terms
of luminosity.

\subsection{Granulation}

We define  $\left ( { {\delta T} /  T } \right
)_g  = \Theta_g  (t,\mu,\phi) $ as the relative temperature perturbations due to the
granulation at the instant $t$ and the position
$(\theta,\phi)$.

As for the mode, we derive the flux perturbation, $\delta
I_{g,\lambda}$,  due to the granulation:
\eqn{
\left ( { {\delta I} \over \bar{I} } \right )_g (t)  =  { { \int_\lambda d\lambda \, E(\lambda)
    \, \delta F_{g,\lambda } \over { \int_\lambda d\lambda \, E(\lambda)
    \, F_\lambda }  } }
}
with
\eqn{
\delta F_{g,\lambda } =   \left ( { {d
    B_\lambda} \over {d \ln T} } \right
) \,  \int_0^{2 \pi}  d \phi \, \int_0^1 d \mu \, \mu \,
\, g_\lambda(\mu) \, \Theta_g  (t,\mu,\phi)
\label{delta_I_g}
}

To go further, one needs to know how
temperature fluctuations due to the granules are distributed along the
star surface. We note that we are only interested in the time averaged
intensity fluctuations.
As a simplification, we assume that distribution of the
temperature fluctuations  is --~ in time average ~-- homogeneous.
Accordingly, we can ignore the dependence of $ \Theta_g $ with
$(\mu,\phi)$.
This is formally equivalent to assume in \eq{dT_T_osc} that $Y_\ell^m
=1 $, as for  a radial mode ($(\ell,m)=(0,0)$).
Then, the expression for $\left ( \delta I / \bar{I} \right
)_{g}$, is derived from \eqs{delta_I_osc_2} and (\ref{R_lm_2})  by
assuming $(\ell,m)=(0,0)$. Accordingly, $\delta
\left ( I / \bar{I} \right )_g$ can be written as
\eqn{
\left ( { {\delta I} \over \bar{I} } \right )_g (t) = R_{g}
\,  \Theta_g   ( t) =  { R_{g} \over 4 } \, \left (\delta L \over
\bar{L} \right )_g  ( t)
\label{delta_I_g_2}
}
with
\eqna{
R_{g}  & =  & R_{\ell=0,m=0}   =   { {\int_\lambda d\lambda \, E(\lambda)
    \,  { {d B_\lambda} \over {d \ln T} }   } \over { \int_\lambda d\lambda \, E(\lambda)
    \, B_\lambda }   } \\
\left (\delta L \over  \bar{L} \right )_g  & = &  4 \, \Theta_g
\label{amp_L_g}
}
As for the radial modes, $\left (\delta L /  \bar{L} \right )_g$
represents the bolometric and instrinsic
luminosity variation due to the granulation.

\end{appendix}

\end{document}